\documentclass[aps,pra,twocolumn,showpacs,showkeys,groupedaddress]{revtex4}
\usepackage{graphicx}
\usepackage{graphics}
\usepackage{xcolor}
\usepackage{ulem}

\newcommand{\ngp}[1]{#1}%\textcolor{red}{#1}}
\newcommand{\etal}{{\it et al.}~}
\newcommand{\etalcc}{{\it et al.}}

\begin{document}
\title{Generation and Decay of Two-Dimensional Quantum Turbulence in a Trapped Bose-Einstein Condensate}
\author{G.~W. Stagg\email{george.stagg@ncl.ac.uk}}
\author{A.~J. Allen}
\author{N.~G. Parker}
\author{C.~F. Barenghi}
\affiliation{Joint Quantum Centre (JQC) Durham-Newcastle, School of Mathematics and Statistics,
Newcastle University, Newcastle upon Tyne NE1 7RU, England, UK.}%

\date{\today}
\begin{abstract}
In a recent experiment, Kwon {\it{et. al}} (arXiv:1403.4658 [cond-mat.quant-gas]) generated a disordered state of quantum vortices by translating an oblate Bose-Einstein condensate past a laser-induced obstacle and studied the subsequent decay of vortex number.   Using mean-field simulations of the Gross-Pitaevskii equation, we shed light on the various stages of the observed dynamics.
We find that the flow of the superfluid past the obstacle leads 
initially to the formation of a classical-like wake, which later becomes 
disordered.  Following removal of the obstacle, the vortex number decays 
due to vortices annihilating and drifting to the boundary. 
Our results are in excellent agreement with the experimental observations.  
Furthermore, we probe thermal effects through phenomenological dissipation.
\end{abstract}
\pacs{03.75.Lm, 03.75.Kk, 67.85.-d, 67.25.dk}
\keywords{vortices, vortex dynamics, quantum turbulence, Bose-Einstein condensates}

\maketitle

\section{Introduction}
Ultracold gaseous Bose-Einstein condensates (BECs) provide a unique testbed with which to investigate the phenomenon of quantum turbulence and the more rudimentary realm of superfluid vortex dynamics \citep{white_anderson_14,barenghi_skrbek_14}.  These systems provide an impressive degree of parameter manipulation unavailable in superfluid helium, the traditional context for studying quantum turbulence \citep{barenghi_donnelly_01}, with scope to control the particle interactions and potential landscape in both time and space.  The typical size of these systems is only one or two orders of magnitude larger than the inter-vortex spacing, which in turn is another order of magnitude larger than the vortex core size.  These compact length scales mean that the collective behaviour of vortices and their interaction with the background condensate is significant.  The emergence of turbulent-like behaviour in the form of a vortex tangle was observed by Henn {\it et al.} in 2009 by oscillating a three-dimensional condensate \cite{henn_seman_09}.  What's more, the experimentalist's handle over the confining potential enables crossover to two-dimensional quantum turbulence~\cite{parker2005}: by tightly confining the trap geometry along one axis, such that the vortices closely embody point vortices \cite{middelkamp}, states of two-dimensional quantum turbulence have been recently reported~\citep{neely_bradley_13,kwon_moon_14}.

In the recent experiment of Kwon {\it et al.} \citep{kwon_moon_14}, a trapped, oblate BEC was translated past a stationary, laser-induced obstacle.  As is characteristic of superfluids, vortices and anti-vortices were nucleated into the condensate once the relative speed exceeded a critical value~\cite{frisch_92}.  \ngp{A state of two-dimensional quantum turbulence emerged, characterized by a disordered distribution of vortices.}  The authors monitored the number of vortices, revealing the dependence on the relative speed and the thermal relaxation of the vortices.  They directly observed vortex collision events, characterized by a crescent-shaped depletion in the condensate density. Furthermore, some vortex cores were seen to coalesce, evidence of vortex pair annihilation.

 In this article we elucidate these experimental
findings through mean-field simulations of the two--dimensional (2D) Gross-Pitaevskii equation (GPE), both at zero-temperature and in the presence of 
thermal dissipation, modelled through a phenomenological dissipation term in 
the GPE.  Notably, our simulations provide insight into the sign of the circulation of the vortices and the early-stage evolution, not accessible experimentally.  We establish the key stages of the dynamics, from the initial nucleation of vortices and formation of a quasi-classical wake, through the rapid symmetry breaking and disorganization of the vortices, to the decay of the vortices by annihilation or passage out of the condensate.  Our approach gives excellent agreement with the experimental observations.  

\section{Set-Up}
In the experiment, a $^{23}$Na condensate with $N=1.8\times 10^6$ atoms was confined within a highly-oblate cylindrically symmetric harmonic trap $V_{\rm{trap}}(x,y,z)=\frac{1}{2}m[\omega_r^2 (x^2+y^2) +\omega_z^2 z^2 ]$, with axial frequency $\omega_z=2 \pi \times 350$ Hz and radial frequency $\omega_r=2\pi \times 15$ Hz (corresponding to an aspect ratio parameter $\omega_z/\omega_r \approx 23$) and where $m$ denotes the atomic mass.  
A 2D mean-field description is strictly valid when 
the condition $N a l_z^3/l_r^3 \ll 1$ is satisfied, 
where $l_z=\sqrt{\hbar/m \omega_z}$ and $l_r=\sqrt{\hbar/m\omega_r}$ 
are the axial and radial harmonic oscillator lengths and $a$ is 
the {\it s}-wave scattering length \cite{delgado,parker2008}.  
For this experiment, $N a l_z^3/l_r^3=8.3$, i.e. the system remains 
3D in nature.   Nonetheless, the dynamics of the vortices is essentially 2D 
because of the suppression of Kelvin waves in 
the $z$-direction~\citep{jackson_proukakis_09}.  
Therefore, we will adopt a 2D description throughout this work and 
show that it is sufficient to capture the experimental observations.  
It is worth noting that in the $xy$ plane the condensate 
closely approximates a Thomas-Fermi (inverted parabola) density 
profile with radius $R_{\rm TF}\approx70 \mu m$.

We parameterize the condensate by a 2D wavefunction $\phi(x,y,t)$; the condensate density distribution follows as $n(x,y,t)=|\phi(x,y,t)|^2$.  The wavefunction satisfies the 2D GPE:
\begin{equation}
%(i-\gamma)
  i\hbar \frac{\partial \phi}{\partial t} = \left(- \frac{\hbar^2}{2m} \nabla ^2  + V(x,y,t) + g |\phi|^2 - \mu \right) \phi
\label{eq:GPE}
\end{equation}
where $\mu$ denotes the chemical potential of the condensate and $g=2 \hbar a (2 \pi \omega_z \hbar/m)^{1/2}$ characterizes the effective 2D nonlinear interactions arising from {\it s}-wave atomic collisions. We solve the GPE on a $1024 \times 1024$ grid using a fourth-order Runge-Kutta method. \ngp{The vortex core size is characterized by the healing length $\xi=\hbar/\sqrt{m n g}$; at the condensate centre this has the value $\xi \approx 0.6 \mu$m.  The grid spacing is $0.27\mu m$ in both $x$ and $y$, and we have verified that reducing the grid spacing has no effect on our results.}  

Following the experiment, the total potential acting on the condensate $V(x,y,t)$ is the above harmonic trap plus a static Gaussian-shaped obstacle potential $V_{\rm{obs}}(x,y) = V_0 \exp{[-{2(x^2 + y^2)}/{d^2}]}$, located at the origin, with $V_0=15 \mu$ and $d=15\mu$m.  The initial ground-state BEC is obtained by solving the GPE in imaginary time with enforced norm $N=1.8\times 10^6$.  At $t=0$ the harmonic trap is centered at $x=18.5\mu$m. The trap is translated towards the left, at speed $v$, over a distance of $37 \mu$m; to smooth this speed curve we additionally include a linear acceleration/deceleration over $3.75$ms at the start/end, which is included as part of the $37\mu$m translation.  Once the trap is at rest, the obstacle amplitude $V_0$ is ramped down to zero over $0.4$s.

\section{Results}
\subsection{Number of Vortices Generated}
\begin{figure}
\includegraphics[width=0.65\linewidth]{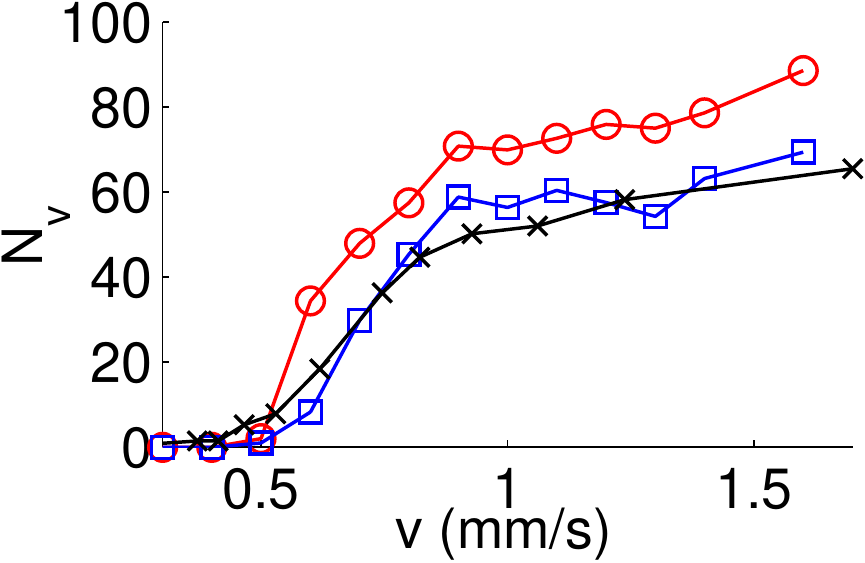}
\caption{\label{fig:N_vV} (Color online) Number of vortices $N_v$ in the condensate after removal of the obstacle. Shown are simulations of the GPE without dissipation (red circles), with dissipation $\gamma = 0.0003$ (blue squares) and experimental results extracted from Fig.~1 of~\citep{kwon_moon_14} (black crosses). Each point is averaged over $20$ ms once the obstacle amplitude reaches $V_0=0$.  For comparison, the speed of sound in the center of the BEC is $v_c\approx 4.6$ mm/s.  }
\end{figure}

Following removal of the obstacle, we determine the number of vortices in the system $N_v$ (performed by identifying locations where the condensate possesses a $2\pi$ singularity in the phase).  
We limit our search to $75$ percent
 of the Thomas-Fermi radius \ngp{(centred on the centre-of-mass to account for sloshing motion)}; by avoiding the low density periphery we avoid artifacts from ghost vortices and match closely what is performed experimentally (since vortices close to the edge are not detected due to low signal-to-noise~\citep{shin_private}).  In Fig. \ref{fig:N_vV} we plot $N_v$ versus the translation speed $v$.  We see the same {\it qualitative} form between our simulations (red circles) and the experiment (black crosses): above a critical speed $v_c \approx 0.45$mm/s vortices enter the system, nucleated by the relative motion between the obstacle and the superfluid, and for $v>v_c$ the growth in $N_v$ is initially rapid but tails off for $v\gg v_c$. Quantitatively, however, the GPE overestimates $N_v$.   One can expect that thermal dissipation, not accounted for in the GPE, will act to reduce the number of vortices in the system.  We introduce the effects of such dissipation via the addition of phenomenological dissipation, $\gamma$~\citep{choi_morgan_98,tsubota_kasamatsu_02}, which enters the GPE~(\ref{eq:GPE}) by replacing $i$ on the left hand side by $(i-\gamma)$.  This term induces the decay of excitations; for single vortices this manifests in them spiraling out of the trapped condensate~\citep{madarassy_barenghi_08,jackson_proukakis_09,allen_zaremba_13,yan_proukakis_14}.  
We choose a small value $\gamma = 0.0003$ so as to model the experiment in its very coldest realization of $\sim130$nK and enforce the norm throughout the dissipative simulations so as to emulate the experiment (for which no significant loss of atom number was observed).

 With this dissipation the data for $N_v$ becomes reduced, bringing it closely in line with the experimental data. Experimental limitations in resolving and counting vortices may also contribute to the over-estimate of $N_v$ from the GPE.

\subsection{Stages of the Condensate Evolution}

We now examine in detail the evolution of the condensate, 
charting its dynamics from the initial stage (when the
harmonic trap translation begins) to the intermediate and final stages 
(randomization and decay of the vortices).  We see the same 
qualitative evolution with and without dissipation, and for all 
velocities exceeding $v_c$.  For the purposes of illustration, 
we focus on an example with dissipation and a translation 
speed $v=1.4$mm/s.  

\begin{figure}
\includegraphics[width=\linewidth]{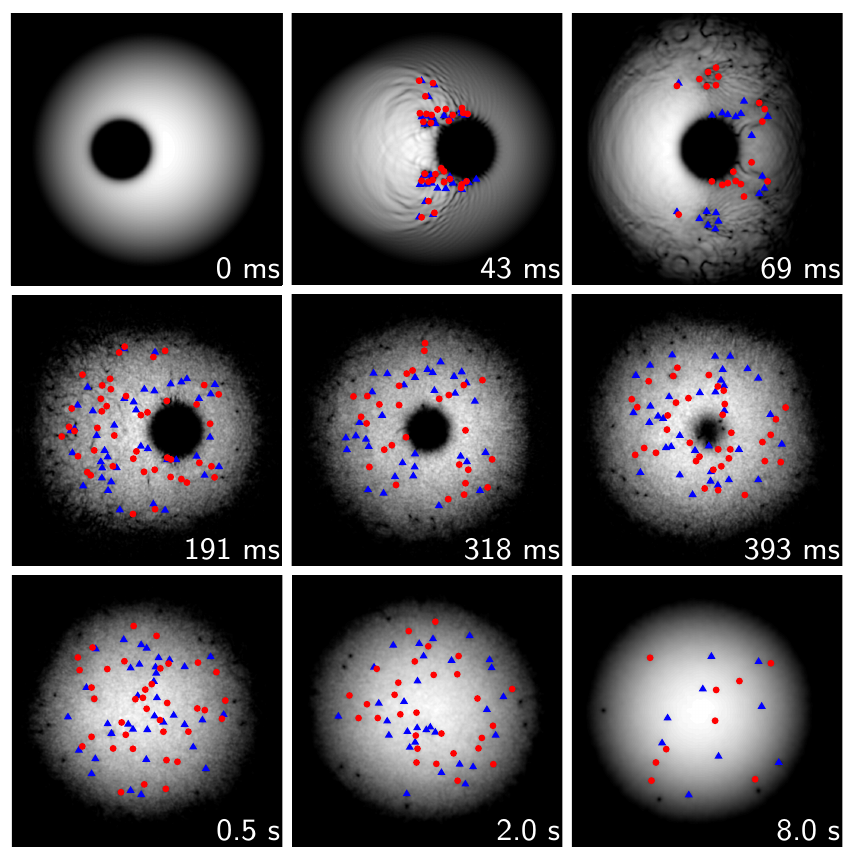}
\caption{\label{fig:densSnapshots} (Color online) Snapshots of the condensate density, for a translational speed $v=1.4$mm/s and in the presence of dissipation ($\gamma=0.0003$). The obstacle is completely removed at $0.43$s. The field of view in each subfigure is of size $[170\mu$m$]^2$ and shifted along the $x$-axis so as to best display the condensate.  Vortices with positive (negative) circulation are highlighted by red circles (blue triangles).
}
\end{figure}

Figure \ref{fig:densSnapshots} shows the condensate density at various times. At the start of the simulation ($t=0$) the condensate has a smooth circular density profile, with a density depression due to the obstacle.  Later vortices appear as small dots of low density; superimposed red/blue markers tag vortices of positive/negative circulation.

\subsubsection{Vortex Nucleation and Wake Formation}

To initiate the dynamics, the harmonic trap is translated to the left.  This is performed sufficiently rapidly that the condensate does not adiabatically follow the trap minimum, but rather begins a sloshing motion in the trap; the centre-of-mass of the BEC oscillates at the trap frequency and the BEC undergoes a quadrupolar shape oscillation.  As the BEC sloshes first to the left, its speed increases.  When the local fluid velocity exceeds the speed of sound \cite{frisch_92}, vortices nucleate at the poles of the obstacle
(where the local fluid velocity is the greatest) 
and are washed downstream (to the left).  
The pattern of vortices nucleated by a moving obstacle 
in a superfluid depends, in general, on the  speed, shape and size of 
the obstacle~\citep{jackson_mccann_00,sasaki_suzuki_10,stagg_parker_14}. 
During the initial evolution vortices of negative
and positive circulation are created near each pole in an 
irregular manner, sometimes with alternating circulation;  
other times several vortices of the same circulation appear.  
In our case, the rate of vortex nucleation is sufficiently 
high that the vortices interact strongly with each other, 
collectively forming macroscopic clusters of negative and positive 
vortices downstream of the object ($t=43$ms).  This is reminiscent of the wakes in classical viscous fluids past cylindrical obstacles \cite{stagg_parker_14}.  
During this early stage, vortices of opposite 
circulation may become very close and annihilate (i.e. undergo 
a 2D reconnection), leaving behind density (sound) waves. The condensate then sloshes to the right; this 
motion not only carries the existing vortices to the opposite 
(right) side of the obstacle but nucleates further vortices. 
As the condensate's sloshing mode is damped by 
the dissipation, the relative speed of the obstacle decreases
and the vortex nucleation pattern changes: 
like-signed vortices are generated near each pole, 
forming symmetric classical--like wakes~\cite{stagg_parker_14}. 
This effect leads to further clustering of like-signed vortices   
($t=69$ms). As the condensate continues to slosh, more
vortices nucleate into the system. It must be stressed that,
up to these early times ($t=191$ms), the vortex distribution remains symmetric 
about the $x$ axis.  \ngp{Without the dissipation term in the GPE, the sloshing mode initially decays while the obstacle is present but then persists with constant amplitude once the obstacle is removed.  If dissipation is included then the sloshing mode continues to decay.} %To counteract this continuous sloshing, our vortex detection subroutine searches around the centre of mass (rather than the center of the trap) in all calculations of $N_v$.}

\subsubsection{Vortex Randomization}
In the presence of the obstacle and the sloshing mode,
vortices continually nucleate and their spatial distribution remains
approximately symmetric about the $x$ axis.  
At later times ($t>318$ms) this symmetry breaks and the vortices 
evolve into a completely disorganised, apparently random 
configuration with no significant clustering of like-signed vortices.  \ngp{This random distribution of vortices is consistent with the experimental observations \cite{kwon_moon_14}; following this we also classify the system as one of quantum turbulence.
Besides vortices, the condensate contains also collective modes and
an energetic, disordered sound field, with this spatial range of excitations further indicative of two-dimensional 
quantum turbulence \cite{parker2005,neely_bradley_13}. (Note that the typical characteristic diagnostics of steady-state 2D quantum turbulence, e.g. power-law energy spectra and the inverse energy cascade, are not appropriate here since the system is not continuously driven and does not reach steady state.)}

\ngp{The vortex randomization is driven by the growth of numerical noise.  We have repeated our results in the presence of imposed noise (amplitude $5\%$, as described elsewhere \cite{stagg_parker_14}) and find the qualitative dynamics to be unchanged (although, as one would expect, the vortex randomization occurs at a slightly earlier time).  This noise serves to model the natural fluctuations that arises in a realistic experimental scenario, e.g. due to thermal and quantum atomic fluctuations, electromagnetic noise, vibrations, etc.}

It is interesting to note the obstacle is still in the system 
at this point, nucleating vortices in a symmetrical manner. 
The disorganised vortices already in the system create a velocity 
field which quickly mixes newly created vortices nucleated at the
poles of the obstacle. Visual inspection, confirmed by a 
clustering-detection algorithm~\citep{white_barenghi_12,reeves_billam_13}, 
shows no significant clusters beyond this stage of the evolution. 
By the time the obstacle is removed the vortex configuration is 
essentially random, but 
the number of positive and negative vortices stays approximately equal.
It is important to remark that, without detecting the sign of the vortex circulation, we
could not reach these conclusions.

\begin{figure}
\includegraphics[width=0.7\linewidth]{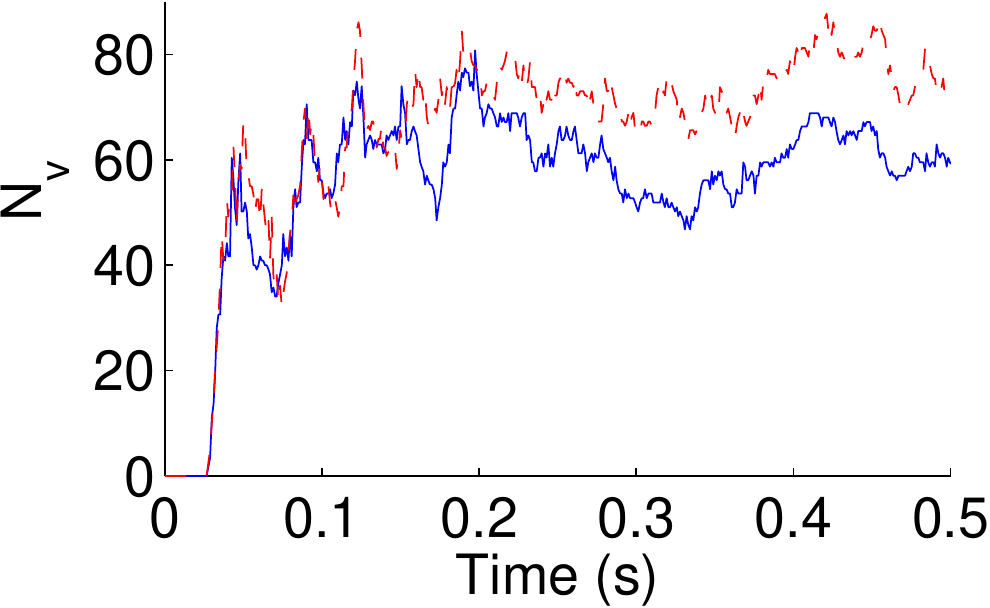}
\caption{\label{fig:N_vTime} (Color online) Growth of vortex number (in a single realization) at early times for a translational speed of $v=1.4$mm/s. Shown are the results with no dissipation (red dashed line) and with dissipation $\gamma=0.0003$ (blue solid line).
}
\end{figure}
\subsubsection{Vortex Decay}
It is clear from Fig. \ref{fig:densSnapshots} that, following the removal 
of the obstacle, the number of vortices ($N_v$) depletes.   
Indeed, one expects that the condensate will decay towards its 
vortex--free, time--independent ground state.  To quantify the vortex generation and
decay, Fig.~\ref{fig:N_vTime} plots $N_v$ versus time. 
The onset of vortex nucleation is at around $t=0.02$ms; 
this is the time taken for accelerating condensate to exceed the speed of 
sound at the poles of the object.
\ngp{At first $N_v$ grows steeply, as vortices (around 40-60) are rapidly driven into the system. Subsequently, $N_v$ grows more slowly; vortices continue to be nucleated from the obstacle but vortices undergo annihilation or move into low density regions where they are not detected.  The fluctuations in $N_v$ are amplified, particularly at early times, by the shape oscillations of the condensate, which carry vortices in and out of the detection radius. As the obstacle is removed at $t\approx 0.4$s, the surrounding condensate fills the low density area. Vortices (including some outside of the detection radius) move inwards with the flow, causing $N_v$ to peak. }
\begin{figure}
\includegraphics[width=1.0\linewidth]{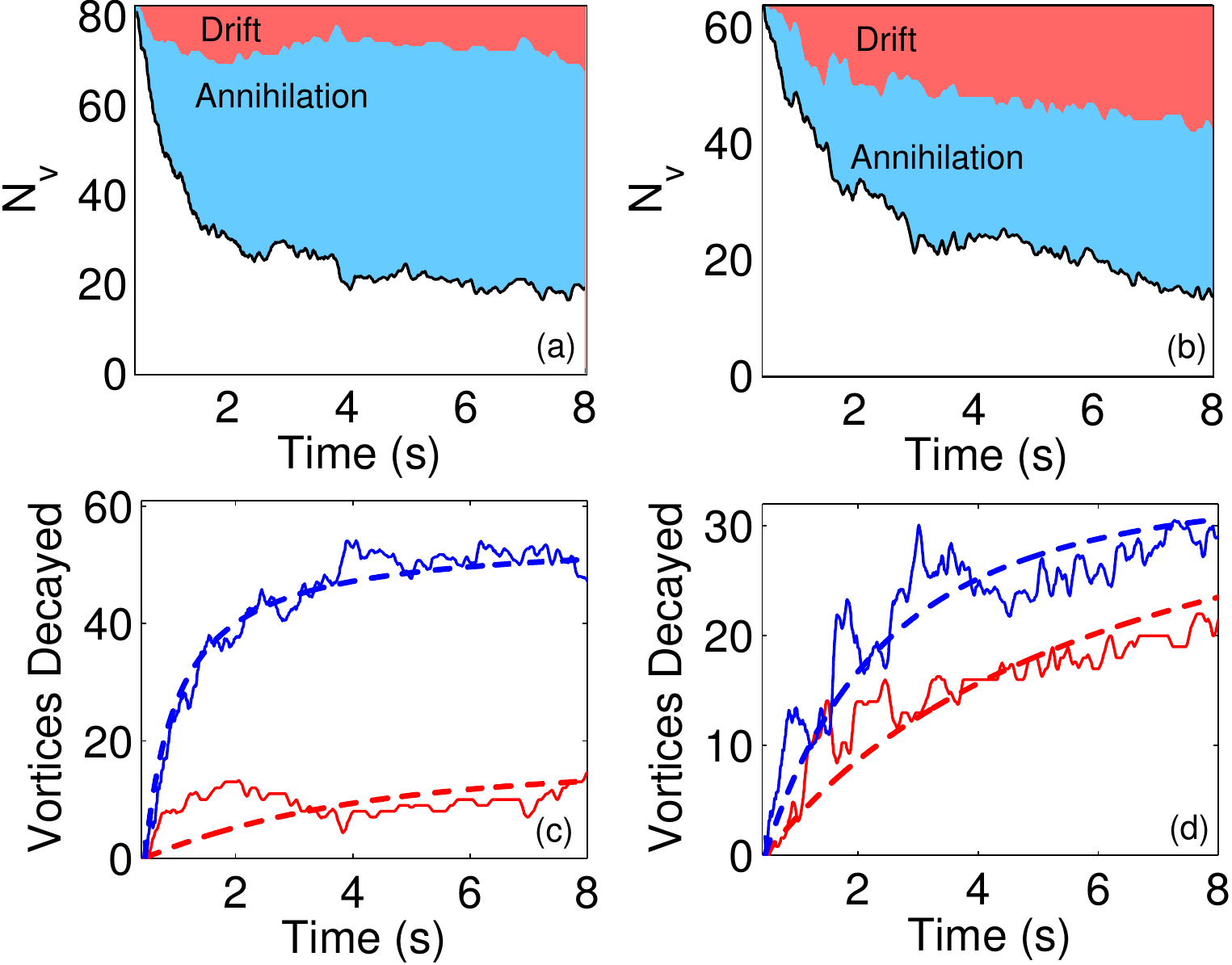}
\caption{\label{fig:N_vLong} (Color online) Vortex decay in the absence of dissipation (a, c) and with dissipation $\gamma=0.0003$ (b, d) for a translational speed of $v=1.4$mm/s.  The upper figures show the decay of the total vortex number $N_v(t)$, with the contribution of drifting and annihilation depicted by the shaded regions.  The lower figures show the drift number $N_d(t)$ and annihilation number $N_a(t)$, plus their respective fits.}
\end{figure}

\ngp{ Following removal of the obstacle, the vortex number $N_v$ decays with time.  This is shown in Fig.~\ref{fig:N_vLong}(a) and (b) for the absence and presence of dissipation, respectively. }  Kwon {\etal}~\citep{kwon_moon_14} argued that there are two mechanisms 
by which vortices decay: (i) thermal dissipation
(resulting in drifting of vortices to the edge of the condensate),
and (ii) vortex-antivortex annihilation events, and proposed
that the vortex decay takes the form:
\begin{equation}
\frac{{\rm d}N_v}{{\rm d}t}= - \Gamma_1 N_v - \Gamma_2 N_v^2.
\label{eqn:N_v_decay}
\end{equation}
Here the linear and nonlinear terms, parameterized by the positive
coefficients $\Gamma_1$ and $\Gamma_2$, respectively, model
these two decay processes.  \ngp{From our simulations we are able to independently count the number of vortices which drift out and the number which annihilate.  
We decompose the number of vortices according to $N_v(t) = N_{v0} - N_d(t) - N_a(t)$, where $N_{v0}$ is the initial number of vortices (when the obstacle is removed), $N_d(t)$ is the cumulative number of vortices which have drifted out of the condensate and $N_a(t)$ is the cumulative number which have undergone pair annihilation. % and $N_c$ is the number which remain within the system (which is finite in the absence of dissipation).
The contribution of both vortex drifting and annihilation to the overall decay of $N_v$ is depicted by the coloured regions in Fig.~\ref{fig:N_vLong}(a) and (b).  In the absence of dissipation  the vortex decay is dominated by annihilation.  Indeed, apart from at early times (where internal condensate dynamics carry vortices out to high radii), no vortices drift out.  In contrast, in the presence of dissipation, vortices continue to drift out over time, consistent with dissipative dynamics of single vortices \citep{allen_zaremba_13}.}
%Our simulations support their findings.

Our decomposition of $N_v$ enables us to independently fit the drift and annihilation decay processes (as two coupled ODEs for $N_d$ and $N_a$, equivalent to Eq. (\ref{eqn:N_v_decay})), with the results shown in Fig. \ref{fig:N_vLong}(c) and (d).  In the absence of dissipation, we find $\Gamma_{2} = 0.0040$.  (It is not appropriate to discuss $\Gamma_1$ since $N_d(t)$ is not of a decaying form) \footnote{While the experimental observations \cite{kwon_moon_14} suggest $\Gamma_2$ is proportional to $T^2$ and thus approaches 0 as $T\rightarrow0$, our results demonstrate a finite $\Gamma_2$ in this limit.}.  In the presence of dissipation we obtain $\Gamma_{1} = 0.093$ and $\Gamma_{2} = 0.0041$, which are comparable to the coldest experiments of Kwon \etal

%\gws{\sout{Fitting our $N_v(t)$ data (with the constraint
%$\Gamma_{1},\Gamma_{2} > 0$) to Eq.~\ref{eqn:N_v_decay},
%we find $\Gamma_{1} = 0$, $\Gamma_{2} = 0.0066$ for the dissipation-free GPE,
%meaning that the only significant decay is through annihilations, 
%and $\Gamma_{1} = 0.113$, $\Gamma_{2} = 0.00439$ in the presence of
%dissipation, which compares well with the coldest experimental data 
%in \citep{kwon_moon_14}}}.

\begin{figure}
\includegraphics[width=0.8\linewidth]{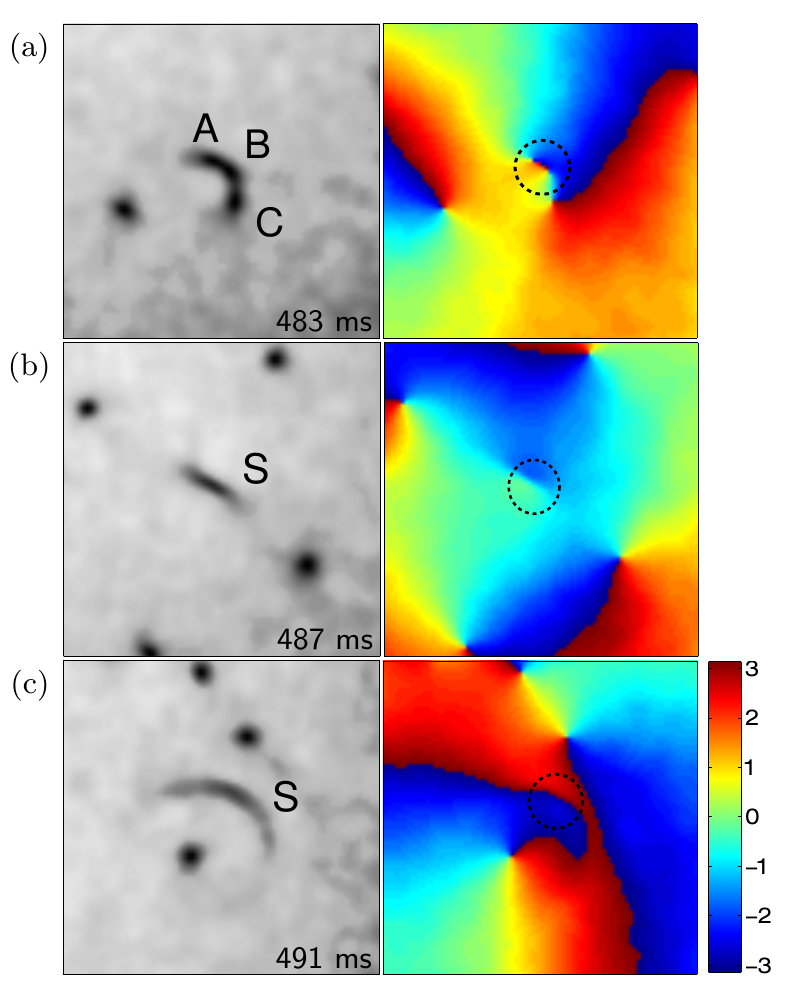}
\caption{\label{fig:cresentPlots} (Color online) Density (left) and phase (right) just before (a), immediately following (b) and a later time after (c) a vortex-antivortex annihilation event.  The field of view is  $[23.5\mu$m]$^2$, centered on the vortex pair/sound pulse (highlighted by a circle in the phase).
}
\end{figure}

\subsection{Crescent-Shaped Density Structures}
In the experiment, Kwon {\etal}observed the occasional appearance of crescent-shaped waves of 
depleted density.  Lacking direct access to the vortex signs,
they suggested that these structures result from
annihilation events of vortices of opposite 
circulation~\citep{nazarenko_onorato_07,rorai_skreenivasan_12,prabhakar_singh_13}: a vortex reconnection is predicted to 
induce an intense, localised, rarefaction 
sound pulse~\cite{leadbeater,zuccher}.  
Figure~\ref{fig:cresentPlots} shows snapshots of the condensate density 
and phase during a reconnection event. Vortices show up as localized dips 
in the density (left column) and $2 \pi$-defects in the phase (right column). 
Figure~\ref{fig:cresentPlots} (a) shows a vortex (A) and antivortex (B) 
close to each other, and a third vortex (C) in the vicinity.
Note that the individual vortices are not spatially resolvable 
through their density alone (the vortex cores merge into a deep, elongated 
crescent-shaped depression), but they are clearly identified by the 
phase plot.  A short time later (b), vortices A and B annihilate, 
as confirmed by the disappearance of their phase singularities, 
leaving behind a shallow rarefaction pulse (S) with a linear phase step.  
This pulse rapidly evolves into a shallow, 
crescent-shaped sound wave [Fig.~\ref{fig:cresentPlots} (c)].  
\ngp{In other words, our simulations yield crescent-shaped density features 
as seen in the experiment, but these features are not uniquely formed 
by annihilation events - they may also result from from two (or more) 
vortices in close proximity.  Information about the condensate phase 
is thus crucial to distinguish the nature
of these observed structures.  In this direction, an approach has recently been proposed for the experimental detection of quantized vortices and their circulation in a 2D BEC \cite{powis}.}

\subsection{Vortex Generation via an Elliptical Obstacle}
\ngp{It is evident from the snapshots in Figure 2 that the initial translation of the condensate past the obstacle generates not just vortices but also shape excitations, sound waves (low-amplitude density waves), and high-amplitude density waves.  These additional excitations will heat the condensate and modify the subsequent turbulent dynamics in a highly nonlinear and complicated manner.  While reducing the translational speed reduces this disruption, this also reduces the number of vortices.  A less disruptive and more efficient (higher rate of vortex nucleation) means to generate vortices may be provided by employing a laser-induced obstacle with {\it elliptical}, rather than circular, cross-section (attainable through cylindrical beam focusing).  Repeating our simulations with such an elliptical obstacle $V_{\rm{obs}}(x,y) = V_0 \exp{[-{2(\epsilon^2 x^2 + y^2)}/{d^2}]}$ with arbitrary ellipticity $\epsilon=3$ (the short/long axis being parallel/perpendicular to the flow) confirms the same qualitative behaviour as for homogeneous systems \citep{stagg_parker_14}: the ellipticity acts to reduce the critical superfluid velocity and, for a given flow speed, increase the rate of vortex nucleation. To illustrate the merits of the elliptical obstacle, in Fig. 6 we depict snapshots of the condensate dynamics for ellipticity $\epsilon=3$ and a translational speed of $v=0.8$mm/s. Despite a lower translational speed, the number of vortices generated by the time the obstacle is removed is almost identical to the circular example of Fig. 3.  As a consequence of the reduced translational speed, the condensate disruption is visibly reduced. It is also worth noting that the elliptical obstacle promotes the formation of clusters of like-signed vortices (see intermediate time), and thus may facilitate future exploration of coherent vortex structures.}

\begin{figure}
\medskip
\includegraphics[width=1\linewidth]{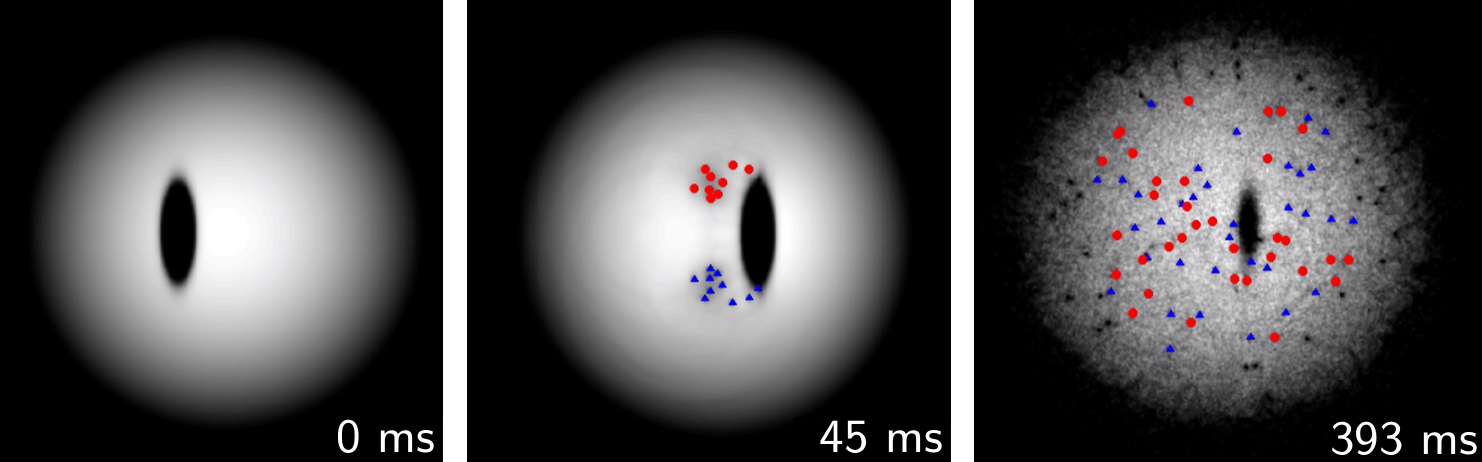}
\caption{\label{fig:ellipse} (Color online) Snapshots of the condensate density for a translational of speed $v=0.8$mm/s past an elliptical obstacle (ellipticity $\epsilon=3$). The field of view in each subfigure is of size $[170\mu$m$]^2$ and shifted along the $x$-axis so as to best display the condensate.  Compared to the corresponding snapshots in Figure 2, the elliptical obstacle generates as many final vortices but at a lower translational speed and with reduced condensate disruption.
}
\end{figure}
\section{Conclusion}
In conclusion, we have shown that the recent experimental creation and decay of vortices within a BEC~\citep{kwon_moon_14} is well described by simulations of the 2D GPE with phenomenological dissipation (despite the 3D nature of the system).  Theoretical access to the condensate phase, and thus the circulation of the vortices, promotes our understanding of the dynamics.  In the early stages of 
translation of the obstacle, a quasi-classical wake of vortices 
forms behind it, before symmetry breaking causes disorganisation 
of the vortices.  After the obstacle is removed, 
the vortices decay in a manner which is both qualitatively and 
quantitatively consistent with the two mechanisms proposed by 
Kwon \etalcc, i.e. loss of vortices at the condensate edge due to thermal dissipation and vortex-antivortex 
annihilation events within the condensate. 
We confirm the occasional appearance of 
crescent-shaped density features, resulting either from the proximity 
of vortex cores or from a sound pulse which follows a 
vortex-antivortex reconnection.  Finally, we propose that a moving {\it elliptical} obstacle may provide a cleaner and more efficient means to generate two-dimensional quantum turbulence.

\begin{acknowledgments}
We thank Y.Shin and his experimental group for many useful discussions.
AJA and CFB acknowledge funding from the EPSRC (Grant number: EP/I019413/1).
This work made use of the facilities of the N8 HPC provided and funded by the N8 consortium and EPSRC.
\end{acknowledgments}


\begin{thebibliography}{20}
\expandafter\ifx\csname natexlab\endcsname\relax\def\natexlab#1{#1}\fi
\expandafter\ifx\csname bibnamefont\endcsname\relax
  \def\bibnamefont#1{#1}\fi
\expandafter\ifx\csname bibfnamefont\endcsname\relax
  \def\bibfnamefont#1{#1}\fi
\expandafter\ifx\csname citenamefont\endcsname\relax
  \def\citenamefont#1{#1}\fi
\expandafter\ifx\csname url\endcsname\relax
  \def\url#1{\texttt{#1}}\fi
\expandafter\ifx\csname urlprefix\endcsname\relax\def\urlprefix{URL }\fi
\providecommand{\bibinfo}[2]{#2}
\providecommand{\eprint}[2][]{\url{#2}}

\bibitem[{\citenamefont{White et~al.}(2014)\citenamefont{White, Anderson, and
  Bagnato}}]{white_anderson_14}
\bibinfo{author}{\bibfnamefont{A.~C.} \bibnamefont{White}},
  \bibinfo{author}{\bibfnamefont{B.~P.} \bibnamefont{Anderson}},
  \bibnamefont{and} \bibinfo{author}{\bibfnamefont{V.~S.}
  \bibnamefont{Bagnato}}, \bibinfo{journal}{Proceedings of the National Academy
  of Sciences}
  \textbf{\bibinfo{volume}{111}}, \bibinfo{pages}{4719} (\bibinfo{year}{2014}).

\bibitem[{\citenamefont{Barenghi et~al.}(2014)\citenamefont{Barenghi, Skrbek,
  and Sreenivasan}}]{barenghi_skrbek_14}
\bibinfo{author}{\bibfnamefont{C.~F.} \bibnamefont{Barenghi}},
  \bibinfo{author}{\bibfnamefont{L.}~\bibnamefont{Skrbek}}, \bibnamefont{and}
  \bibinfo{author}{\bibfnamefont{K.~R.} \bibnamefont{Sreenivasan}},
  \bibinfo{journal}{Proceedings of the National Academy of Sciences}
  \textbf{\bibinfo{volume}{111}}, \bibinfo{pages}{4647} (\bibinfo{year}{2014}).

\bibitem[{\citenamefont{Barenghi et~al.}(2001)\citenamefont{Barenghi, Donnelly,
  Vinen, and eds.}}]{barenghi_donnelly_01}
\bibinfo{author}{\bibfnamefont{C.}~\bibnamefont{Barenghi}},
  \bibinfo{author}{\bibfnamefont{R.}~\bibnamefont{Donnelly}},
  \bibinfo{author}{\bibfnamefont{W.}~\bibnamefont{Vinen}}, \bibnamefont{and}
  \bibinfo{author}{\bibnamefont{eds.}}, \emph{\bibinfo{title}{Quantized Vortex
  Dynamics and Superfluid Turbulence}} (\bibinfo{publisher}{Springer, Berlin},
  \bibinfo{year}{2001}).

\bibitem[{\citenamefont{Henn et~al.}(2009)\citenamefont{Henn, Seman, Roati,
  Magalh\~aes, and Bagnato}}]{henn_seman_09}
\bibinfo{author}{\bibfnamefont{E.~A.~L.} \bibnamefont{Henn}},
  \bibinfo{author}{\bibfnamefont{J.~A.} \bibnamefont{Seman}},
  \bibinfo{author}{\bibfnamefont{G.}~\bibnamefont{Roati}},
  \bibinfo{author}{\bibfnamefont{K.~M.~F.} \bibnamefont{Magalh\~aes}},
  \bibnamefont{and} \bibinfo{author}{\bibfnamefont{V.~S.}
  \bibnamefont{Bagnato}}, \bibinfo{journal}{Phys. Rev. Lett.}
  \textbf{\bibinfo{volume}{103}}, \bibinfo{pages}{045301}
  (\bibinfo{year}{2009}).
  
  \bibitem{parker2005}  N. G. Parker and C. S. Adams, Phys. Rev. Lett. {\bf 95}, 145301 (2005).

\bibitem[{\citenamefont{Middelkamp et~al.}(2011)\citenamefont{Middelkamp,
  Torres, Kevrekidis, Frantzeskakis, Carretero-Gonz\'alez, Schmelcher,
  Freilich, and Hall}}]{middelkamp}
\bibinfo{author}{\bibfnamefont{S.}~\bibnamefont{Middelkamp}},
  \bibinfo{author}{\bibfnamefont{P.~J.} \bibnamefont{Torres}},
  \bibinfo{author}{\bibfnamefont{P.~G.} \bibnamefont{Kevrekidis}},
  \bibinfo{author}{\bibfnamefont{D.~J.} \bibnamefont{Frantzeskakis}},
  \bibinfo{author}{\bibfnamefont{R.}~\bibnamefont{Carretero-Gonz\'alez}},
  \bibinfo{author}{\bibfnamefont{P.}~\bibnamefont{Schmelcher}},
  \bibinfo{author}{\bibfnamefont{D.~V.} \bibnamefont{Freilich}},
  \bibnamefont{and} \bibinfo{author}{\bibfnamefont{D.~S.} \bibnamefont{Hall}},
  \bibinfo{journal}{Phys. Rev. A} \textbf{\bibinfo{volume}{84}},
  \bibinfo{pages}{011605} (\bibinfo{year}{2011}).

\bibitem[{\citenamefont{Neely et~al.}(2013)\citenamefont{Neely, Bradley,
  Samson, Rooney, Wright, Law, Carretero-Gonz\'alez, Kevrekidis, Davis, and
  Anderson}}]{neely_bradley_13}
\bibinfo{author}{\bibfnamefont{T.~W.} \bibnamefont{Neely}},
  \bibinfo{author}{\bibfnamefont{A.~S.} \bibnamefont{Bradley}},
  \bibinfo{author}{\bibfnamefont{E.~C.} \bibnamefont{Samson}},
  \bibinfo{author}{\bibfnamefont{S.~J.} \bibnamefont{Rooney}},
  \bibinfo{author}{\bibfnamefont{E.~M.} \bibnamefont{Wright}},
  \bibinfo{author}{\bibfnamefont{K.~J.~H.} \bibnamefont{Law}},
  \bibinfo{author}{\bibfnamefont{R.}~\bibnamefont{Carretero-Gonz\'alez}},
  \bibinfo{author}{\bibfnamefont{P.~G.} \bibnamefont{Kevrekidis}},
  \bibinfo{author}{\bibfnamefont{M.~J.} \bibnamefont{Davis}}, \bibnamefont{and}
  \bibinfo{author}{\bibfnamefont{B.~P.} \bibnamefont{Anderson}},
  \bibinfo{journal}{Phys. Rev. Lett.} \textbf{\bibinfo{volume}{111}},
  \bibinfo{pages}{235301} (\bibinfo{year}{2013}).

\bibitem[{\citenamefont{Kwon et~al.}(2014)\citenamefont{Kwon, Moon, Choi, Seo,
  and Shin}}]{kwon_moon_14}
\bibinfo{author}{\bibfnamefont{W.~J.} \bibnamefont{Kwon}},
  \bibinfo{author}{\bibfnamefont{G.}~\bibnamefont{Moon}},
  \bibinfo{author}{\bibfnamefont{J.}~\bibnamefont{Choi}},
  \bibinfo{author}{\bibfnamefont{S.~W.} \bibnamefont{Seo}}, \bibnamefont{and}
  \bibinfo{author}{\bibfnamefont{Y.}~\bibnamefont{Shin}}
  \bibinfo{note}{arXiv:1403.4658 [cond-mat.quant-gas]},
  \eprint{1403.4658} (\bibinfo{year}{2014}).

\bibitem[{\citenamefont{Frisch et~al.}(1992)\citenamefont{Frisch, Pomeau, and Rica}}]{frisch_92}
\bibinfo{author}{\bibfnamefont{T.} \bibnamefont{Frisch}},
  \bibinfo{author}{\bibfnamefont{Y.} \bibnamefont{Pomeau}},
  \bibinfo{author}{\bibfnamefont{S.} \bibnamefont{Rica}},
  \bibinfo{journal}{Phys. Rev. Lett.} \textbf{\bibinfo{volume}{69}},
  \bibinfo{pages}{1644} (\bibinfo{year}{1992}).


\bibitem[{\citenamefont{Mu\~noz Mateo and Delgado}(2006)}]{delgado}
\bibinfo{author}{\bibfnamefont{A.}~\bibnamefont{Mu\~noz Mateo}}
  \bibnamefont{and} \bibinfo{author}{\bibfnamefont{V.}~\bibnamefont{Delgado}},
  \bibinfo{journal}{Phys. Rev. A} \textbf{\bibinfo{volume}{74}},
  \bibinfo{pages}{065602} (\bibinfo{year}{2006}).

\bibitem[{\citenamefont{Parker and O'Dell}(2008)}]{parker2008}
\bibinfo{author}{\bibfnamefont{N.~G.} \bibnamefont{Parker}} \bibnamefont{and}
  \bibinfo{author}{\bibfnamefont{D.~H.~J.} \bibnamefont{O'Dell}},
  \bibinfo{journal}{Phys. Rev. A} \textbf{\bibinfo{volume}{78}},
  \bibinfo{pages}{041601} (\bibinfo{year}{2008}).

\bibitem[{\citenamefont{Jackson et~al.}(2009)\citenamefont{Jackson, Proukakis,
  Barenghi, and Zaremba}}]{jackson_proukakis_09}
\bibinfo{author}{\bibfnamefont{B.}~\bibnamefont{Jackson}},
  \bibinfo{author}{\bibfnamefont{N.~P.} \bibnamefont{Proukakis}},
  \bibinfo{author}{\bibfnamefont{C.~F.} \bibnamefont{Barenghi}},
  \bibnamefont{and} \bibinfo{author}{\bibfnamefont{E.}~\bibnamefont{Zaremba}},
  \bibinfo{journal}{Phys. Rev. A} \textbf{\bibinfo{volume}{79}},
  \bibinfo{pages}{053615} (\bibinfo{year}{2009}).

\bibitem[{\citenamefont{Shin}()}]{shin_private}
\bibinfo{author}{\bibfnamefont{Y.}~\bibnamefont{Shin}},
  \emph{\bibinfo{title}{Private communication}}.

\bibitem[{\citenamefont{Choi et~al.}(1998)\citenamefont{Choi, Morgan, and
  Burnett}}]{choi_morgan_98}
\bibinfo{author}{\bibfnamefont{S.}~\bibnamefont{Choi}},
  \bibinfo{author}{\bibfnamefont{S.~A.} \bibnamefont{Morgan}},
  \bibnamefont{and} \bibinfo{author}{\bibfnamefont{K.}~\bibnamefont{Burnett}},
  \bibinfo{journal}{Phys. Rev. A} \textbf{\bibinfo{volume}{57}},
  \bibinfo{pages}{4057} (\bibinfo{year}{1998}).

\bibitem[{\citenamefont{Tsubota et~al.}(2002)\citenamefont{Tsubota, Kasamatsu,
  and Ueda}}]{tsubota_kasamatsu_02}
\bibinfo{author}{\bibfnamefont{M.}~\bibnamefont{Tsubota}},
  \bibinfo{author}{\bibfnamefont{K.}~\bibnamefont{Kasamatsu}},
  \bibnamefont{and} \bibinfo{author}{\bibfnamefont{M.}~\bibnamefont{Ueda}},
  \bibinfo{journal}{Phys. Rev. A} \textbf{\bibinfo{volume}{65}},
  \bibinfo{pages}{023603} (\bibinfo{year}{2002}).

\bibitem[{\citenamefont{Madarassy and Barenghi}(2008)}]{madarassy_barenghi_08}
\bibinfo{author}{\bibfnamefont{E.}~\bibnamefont{Madarassy}} \bibnamefont{and}
  \bibinfo{author}{\bibfnamefont{C.}~\bibnamefont{Barenghi}},
  \bibinfo{journal}{Journal of Low Temperature Physics}
  \textbf{\bibinfo{volume}{152}}, \bibinfo{pages}{122} (\bibinfo{year}{2008}),
  ISSN \bibinfo{issn}{0022-2291}.


\bibitem{allen_zaremba_13}  A. J. Allen, E. Zaremba, C. F. Barenghi and N. P. Proukakis, Phys. Rev. A {\bf{87}}, 013630 (2013).

\bibitem[{\citenamefont{Yan et~al.}(2009)\citenamefont{Yan, Carretero-Gonz\'alez,
  Frantzeskakis, Kevrekidis, Proukakis, and Spirn}}]{yan_proukakis_14}
  \bibinfo{author}{\bibfnamefont{D.}~\bibnamefont{Yan}},
  \bibinfo{author}{\bibfnamefont{R.} \bibnamefont{Carretero-Gonz\'alez}},
  \bibinfo{author}{\bibfnamefont{D.~J.} \bibnamefont{Frantzeskakis}},
  \bibinfo{author}{\bibfnamefont{P.~G.} \bibnamefont{Kevrekidis}},
  \bibinfo{author}{\bibfnamefont{N.~P.} \bibnamefont{Proukakis}},
  \bibnamefont{and} \bibinfo{author}{\bibfnamefont{D.}~\bibnamefont{Spirn}},
  \bibinfo{journal}{Phys. Rev. A} \textbf{\bibinfo{volume}{89}},
  \bibinfo{pages}{043613} (\bibinfo{year}{2014}).


\bibitem[{\citenamefont{Jackson et~al.}(2000)\citenamefont{Jackson, McCann, and
  Adams}}]{jackson_mccann_00}
\bibinfo{author}{\bibfnamefont{B.}~\bibnamefont{Jackson}},
  \bibinfo{author}{\bibfnamefont{J.~F.} \bibnamefont{McCann}},
  \bibnamefont{and} \bibinfo{author}{\bibfnamefont{C.~S.} \bibnamefont{Adams}},
  \bibinfo{journal}{Phys. Rev. A} \textbf{\bibinfo{volume}{61}},
  \bibinfo{pages}{051603} (\bibinfo{year}{2000}).

\bibitem[{\citenamefont{Sasaki et~al.}(2010)\citenamefont{Sasaki, Suzuki, and
  Saito}}]{sasaki_suzuki_10}
\bibinfo{author}{\bibfnamefont{K.}~\bibnamefont{Sasaki}},
  \bibinfo{author}{\bibfnamefont{N.}~\bibnamefont{Suzuki}}, \bibnamefont{and}
  \bibinfo{author}{\bibfnamefont{H.}~\bibnamefont{Saito}},
  \bibinfo{journal}{Phys. Rev. Lett.} \textbf{\bibinfo{volume}{104}},
  \bibinfo{pages}{150404} (\bibinfo{year}{2010}).

\bibitem[{\citenamefont{Stagg et~al.}(2014)\citenamefont{Stagg, Parker, and
  Barenghi}}]{stagg_parker_14}
\bibinfo{author}{\bibfnamefont{G.~W.} \bibnamefont{Stagg}},
  \bibinfo{author}{\bibfnamefont{N.~G.} \bibnamefont{Parker}},
  \bibnamefont{and} \bibinfo{author}{\bibfnamefont{C.~F.}
  \bibnamefont{Barenghi}}, \bibinfo{journal}{Journal of Physics B: Atomic,
  Molecular and Optical Physics} \textbf{\bibinfo{volume}{47}},
  \bibinfo{pages}{095304} (\bibinfo{year}{2014}); 
  G. W. Stagg, A. J. Allen, C. F. Barenghi and N. G. Parker, arXiv:1411.4842 [cond-mat.quant-gas], 1411.4842 (2014).


\bibitem[{\citenamefont{White et~al.}(2012)\citenamefont{White, Barenghi, and
  Proukakis}}]{white_barenghi_12}
\bibinfo{author}{\bibfnamefont{A.~C.} \bibnamefont{White}},
  \bibinfo{author}{\bibfnamefont{C.~F.} \bibnamefont{Barenghi}},
  \bibnamefont{and} \bibinfo{author}{\bibfnamefont{N.~P.}
  \bibnamefont{Proukakis}}, \bibinfo{journal}{Phys. Rev. A}
  \textbf{\bibinfo{volume}{86}}, \bibinfo{pages}{013635}
  (\bibinfo{year}{2012}).

\bibitem[{\citenamefont{Reeves et~al.}(2013)\citenamefont{Reeves, Billam,
  Anderson, and Bradley}}]{reeves_billam_13}
\bibinfo{author}{\bibfnamefont{M.~T.} \bibnamefont{Reeves}},
  \bibinfo{author}{\bibfnamefont{T.~P.} \bibnamefont{Billam}},
  \bibinfo{author}{\bibfnamefont{B.~P.} \bibnamefont{Anderson}},
  \bibnamefont{and} \bibinfo{author}{\bibfnamefont{A.~S.}
  \bibnamefont{Bradley}}, \bibinfo{journal}{Phys. Rev. Lett.}
  \textbf{\bibinfo{volume}{110}}, \bibinfo{pages}{104501}
  (\bibinfo{year}{2013}).


\bibitem[{\citenamefont{Nazarenko and Onorato}(2007)}]{nazarenko_onorato_07}
\bibinfo{author}{\bibfnamefont{S.}~\bibnamefont{Nazarenko}} \bibnamefont{and}
  \bibinfo{author}{\bibfnamefont{M.}~\bibnamefont{Onorato}},
  \bibinfo{journal}{Journal of Low Temperature Physics}
  \textbf{\bibinfo{volume}{146}}, \bibinfo{pages}{31} (\bibinfo{year}{2007}),
  ISSN \bibinfo{issn}{0022-2291}.

\bibitem{rorai_skreenivasan_12}
C. Rorai, K. R. Sreenivasan and M. E. Fisher, 
Phys. Rev. B {\bf 88}, 134522 (2013).


\bibitem[{\citenamefont{Prabhakar et~al.}(2013)\citenamefont{Prabhakar, Singh,
  Gautam, and Angom}}]{prabhakar_singh_13}
\bibinfo{author}{\bibfnamefont{S.}~\bibnamefont{Prabhakar}},
  \bibinfo{author}{\bibfnamefont{R.~P.} \bibnamefont{Singh}},
  \bibinfo{author}{\bibfnamefont{S.}~\bibnamefont{Gautam}}, \bibnamefont{and}
  \bibinfo{author}{\bibfnamefont{D.}~\bibnamefont{Angom}},
  \bibinfo{journal}{Journal of Physics B: Atomic, Molecular and Optical
  Physics} \textbf{\bibinfo{volume}{46}}, \bibinfo{pages}{125302}
  (\bibinfo{year}{2013}).

\bibitem{leadbeater}
M. Leadbeater, T. Winiecki, D.C. Samuels, C.F. Barenghi, and C.S. Adams,
Phys. Rev. Lett. {\bf 86}, 1410 (2001).
\bibitem{zuccher}
S. Zuccher, M. Caliari, A.W. Baggaley, and C.F. Barenghi,
Phys. of Fluids {\bf 24}, 125108 (2012).
\bibitem{powis}
A. T. Powis, S. J. Sammut and T. P. Simula,
Phys. Rev. Lett. {\bf 113}, 165303 (2014).

\end{thebibliography}
\end{document}